\documentclass[twocolumn,showpacs,preprintnumbers,amsmath,amssymb]{revtex4}
\usepackage{graphicx}
\usepackage{dcolumn}
\usepackage{bm}

\begin{document}
\draft

\title{Development of a deformation-tunable quadrupolar microcavity}

\author{Juhee Yang, Songky Moon, Sang-Bum Lee}
\address{School of Physics, Seoul National University, Seoul 151-747, Korea}

\author{Jeong-Bo Shim}
\address{Department of Physics, Korea Advanced Institute of
Science and Technology, Taejon 305-701, Korea}

\author{Sang-Wook Kim}
\address{Department of Physics Education, Pusan National University, Pusan 609-735, Korea}

\author{Hai-Woong Lee}
\address{Department of Physics, Korea Advanced Institute of
Science and Technology, Taejon 305-701, Korea}

\author{Jai-Hyung Lee, Kyungwon An}
\email{kwan@phys.snu.ac.kr}
\address{School of Physics, Seoul National University, Seoul 151-747, Korea}

\date{\today}

\begin{abstract}
We have developed a technique for realizing a two-dimensional quadrupolar microcavity with its deformation variable from 0\% to 20\% continuously. We employed a microjet ejected from a noncircular orifice in order to generate a stationary column with modulated quadrupolar deformation in its cross section. Wavelength red shifts of low-order cavity modes due to shape deformation were measured and were found to be in good agreement with the wave calculation for the same deformation, indicating the observed deformation is quadrupolar in nature.
\end{abstract}

\pacs{05.45.Mt, 42.55.Sa}

\maketitle

\section{Introduction}
Unlike circular microcavities showing isotropic output emission, asymmetric resonant cavities (ARC's)\cite{Nockel94, Mekis95,Chang96} provide directional output  along with fairly high cavity quality factors up to certain degree of deformation. In addition, the internal ray dynamics of ARC's exhibits an evolution from regularity to chaos as the degree of deformation increases. With these properties, ARC's serve not only as a useful resonators for optoelectronics devices
but also as a versatile tool for studying ray and wave chaos \cite{Nockel97}.

In ARC's, the internal ray dynamics strongly depend on the degree of deformation, and so do mode properties such as output directionality \cite{Gmachl98} and cavity quality factor. Whispering gallery modes and their ultrahigh cavity qualtiy $Q$ factors of circular microcavities can be associated with regular ray dynamics in the cavities. As the deformation is increased gradually, this dynamical regularity begins to be broken following Kolmogorov-Arnold-Moser(KAM) scenario in phase space. The emission pattern also experiences a dramatic change, from isotropic emission into collimated tangential emission at high-curvature points on the surface. When the deformation is further increased, even KAM tori are broken and the internal ray dynamics becomes more chaotic. However, this ray picture is not fully applicable to typical microcavities with their radial size ranging from several microns to tens of microns. In this region, it is not well understood how the chaotic ray dynamics is related to both actual modes of cavity and output emission patterns of these modes.

If one can tune the deformation of a microcavity from a perfect circle to a final deformed form continuously, one can follow the evolution of modes and thus can easily identify the origin and mode numbers of observed modes of ARC. Such identification allows us to perform direct comparison with wave calculation and help us to obtain better understanding on the connection between the mode distribution, cavity quality factors and output directionality of these modes and the aforementioned chaotic ray dynamics.

Notwithstanding the advantage of electrical pumping capability, microcavities made of semiconductor \cite{Gmachl98,Gianordoli00,Gmachl02,Rex02}, are rigid under room temperature and thus cannot be changed in shape continuously. In order to have varied shapes, separate cavities with different deformation should be employed. However, it is almost impossible to make connections among observed modes from the separate cavities. 
In addition, the cavity quality factors are limited to $10^4\sim 10^5$ without cumbersome annealing process even for fairly large microcavities with diameters of hundreds of microns. For such large cavity sizes the free spectral range of modes are comparable to or larger than the linewidth of modes and thus individual modes are difficult to be resolved.

On the other hand, surface-tension-induced microcavities (STIM's), such as liquid droplets \cite{Qian86, Campillo91}, liquid microjet \cite{Lee02}, silica micro spheres \cite{Vernooy98,Moon2003} and microtoroid \cite{Vahala03} can provide ultrahigh-$Q$ factors at relatively small sizes due to their surface smoothness. STIM's have been employed in studying shape-deformation effects \cite{Mekis95,Lacey03,Lee02}. Among various STIM's, liquid microjet technique is particularly suitable for generating deformed two-dimensional microcavities.

In the present work, we have realized a stable stationary microjet ARC with continuously variable deformation by adapting the well-developed droplet generation technique \cite{Berglund73, Lin90,Moon95} and precisely controlling the
hydrodynamics of the microjet. Our liquid microcavity offers another benefit that the concentration of gain molecules for lasing and other spectroscopic measurements can be easily adjusted. Although it is only subject to optical pumping, the liquid microjet ARC is nevertheless a powerful system for studying the aforementioned ray-wave correspondence or
the quantum chaos in the microcavity.

This paper is organized in the following way. In Sec.\ II, we describe a method to realize deformed microcavities with a set of discrete degrees of deformation by using a stationary surface-modulated microjet column. A method to obtain quadrupole cavities with continuously tunable deformation by controlling the pressure of the jet is discussed in Sec.\ III. Spectral analysis of cavity modes in deformed microcavities is presented in Sec.\ IV, where we demonstrate that the wavelength red shift due to shape deformation only can be extracted from raw experimental data. The observed shifts vs.\ the measured degrees of deformation are shown to be in good agreement with the results of wave calculation. 

\section{Method to obtain discrete degree of deformation}
An ARC can be obtained from a horizontal cross section of a
microjet, which is made of ethanol ($n$=1.361) doped with
Rhodamine B dye as gain molecules. A similar microjet generator
with a circular orifice was introduced previously \cite{Moon95}.
However, in the present study the microjet employs a noncircular
orifice as a nozzle.

\begin{figure}
\includegraphics[width=3.4in]{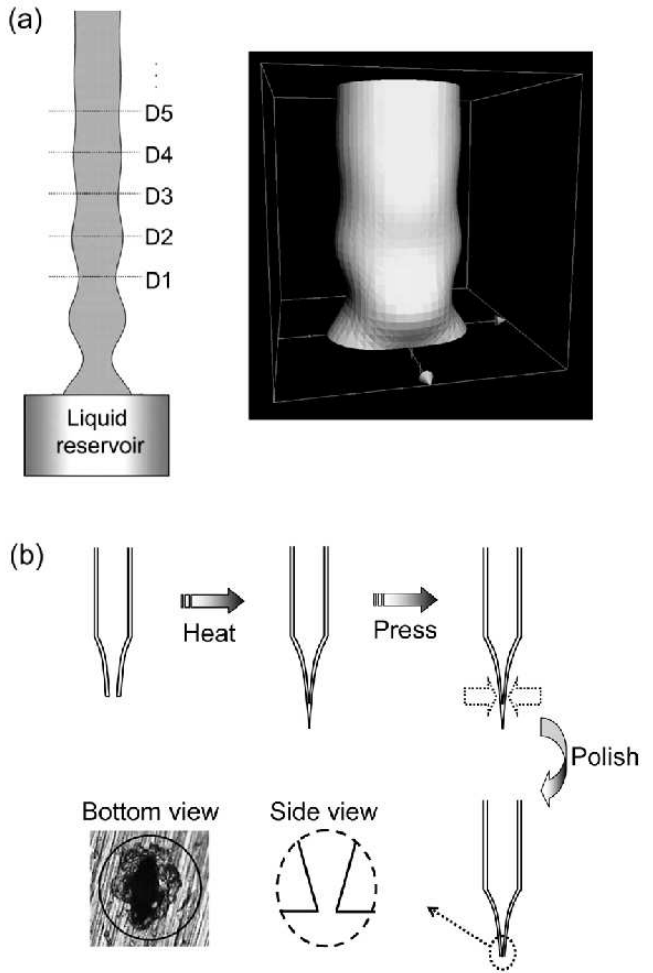}
\caption{(a) Model for the deformed microjet column. (b) Fabrication procedure of a noncircular orifice. The side view shows that the inner walls are inclined. The bottom view is a real image of a noncircular orifice.}
\label{fig1}
\end{figure}

We have reported elsewhere \cite{Lee06} that such a noncircular orifice induces a microjet to form a stationary tidal column as depicted in Fig.\ \ref{fig1}(a). Viscosity of the liquid makes the oscillation amplitude of the column rapidly damped along the jet propagation direction. The shape of the jet column can be approximated in the cylindrical coordinates by the following time-independent equation
\begin{equation}
r(\theta,z)=a\left[1+\eta_0 \exp \left(- \frac{z}{v_z \tau}\right)\sin\left(\frac{2\pi}{v_{z}T}z+\xi\right)\cos2\theta\right]\;,
\label{eq1}
\end{equation}
where $a$ is the mean radius and $\eta_0$ is a seed deformation parameter, $\tau$ is the decay rate, $T$  is the period of
oscillation and $\xi$ is an initial phase of oscillation. Both the decay rate and the period of oscillation are approximately
constant depending on the properties of the liquid and the orifice size. The jet velocity $v_z$ is assumed to be uniform
across the entire jet \cite{Lamb45}.

A two-dimensional ARC, specifically a quadrupole-deformed microcavity (QDM), is obtained by selecting cross-sectional
planes located at the extrema $z_n$'s ($n$=-1, 0, 1, 2,$\ldots$ in Fig.\ \ref{fig1}(a)) of the amplitude oscillation, where the
cavity boundary is given by
\begin{equation}
r(\theta,z_n)=a\left[1+(-)^n\eta_n \cos2\theta\right]\;,
\label{eq2}
\end{equation}
where
\begin{equation}
\eta_n=\eta_0 \exp\left(-\frac{z_n}{v_z \tau}\right)\;.
\label{eq3}
\end{equation}
It is our convention to denote the cross-sectional plane or anti-nodal plane at $z_n$ as D$n$ ($n=1,2,3\ldots$). According to our convention, D1 corresponds to the second minimum of amplitude oscillation from the nozzle in Fig.\ \ref{fig1}(a). Equation (\ref{eq1}) describes a quadrupole of deformation $\eta=\eta_n$, and thus we can obtain QDM's
at D1, D2, D3, etc. The major axis of QDM's at D1, D3, D5, etc is rotated by 90$^\circ$ with respect to that of QDM's at D2, D4, D6, etc.

Degree of possible three-dimensional (3D) effect due to a finite thickness of the region to be used in experiment around the
cross-sectional plane, which is about 10 microns, can be estimated in the following way. Under our experimental conditions to be described below, a typical value of $v_z T$ is 280 microns, which is much larger than the mean radius $a$ of 15 microns.  The possible variation in the cavity size over the 10 $\mu$m region in the $z$ direction is then estimated to be less than 0.1\% or 0.02 micron in diameter, and therefore, the 3D effect can be safely neglected.

The degree of deformation of QDM's can be determined from their diffraction patterns made by an incident probe laser as described in Refs.\ \cite{Lee02} and \cite{Lee06}. Details of diffraction pattern measurement and data analysis are found in Ref.\ \cite{Lee06}. The degrees of deformation of QDM's are found to be 32\% (D1), 19.2\% (D2), 13.3\% (D3), 10.2\% (D4), 6.5\% (D5), etc under 1.8 bar of ejection pressure.

Let us now describe the fabricating process of the noncircular orifice, a key component in our system. Figure \ref{fig1}(b)
depicts the fabrication procedure. A tip of pyrex tube is placed vertically in the center of a Nichrome heating coil. As the tube
melts down, the bottom part of the tube droops. At that moment, the inner walls of the tube are inclined, resembling the letter 'V' as depicted in the side view of Fig.\ \ref{fig1}(b). While being heated, the both sides of the tip are pressed with a tweezer. After slightly grinding off and polishing the tip we can obtain a noncircular orifice on it. The diameter can be controlled from several microns to hundreds of microns by amount of polishing.

\section{Method to obtain continuously tuned degree of deformation}
According to Eq.\ (1), the microjet boundary at the nozzle ($z=0$) is given by
\begin{equation}
r(\theta,0)=a \left(1+\eta_0 \sin \xi \cos 2\theta \right) \;,
\label{eq4}
\end{equation}
which should match the boundary of the orifice. In fact, the boundary of the orifice usually contains higher-order components of small magnitude. These higher-order components have larger damping rates than that of the quadrupole, and therefore, by the time of the jet reaches D1 location, the microjet surface becomes well approximated by Eq.\ (\ref{eq1}).

\begin{figure}
\includegraphics[width=3.4in]{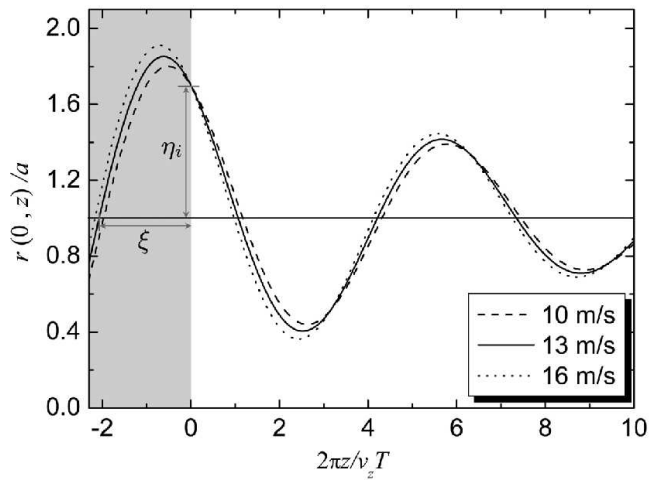}
\caption{Cross sectional view of the deformation oscillation along the $z$ axis, given by Eq.\  (\ref{eq1}) with $\theta=0$. Initial deformation $\eta_i(=\eta_0\sin\xi)$ at $z=0$ is predetermined by the orifice. As the jet velocity $v_z$ is increased,
the initial phase $\xi$ changes in such a way that $\cos\xi$ is proportional to the velocity (see Eq.\ (\ref{eq7})).
Consequently, the deformations at anti-nodal planes, D1, D2, D3 $\ldots$, can be increased as the jet velocity.
The parameter values used for these plots are $T=28 \mu$s, $\tau=39 \mu$s, $a=15 \mu$m, $K$=0.1 and $\eta_i=0.7$.}
\label{amspectrum}
\end{figure}

Since the degree of deformation at the nozzle is more or less prescribed by the shape of the orifice, $\eta_i\equiv \eta_0
\sin\xi$ can be regarded to be fixed. However, $\eta_0$, called a seed deformation, can still be changed under this constraint, and therefore, the deformation $\eta_n$ at anti-nodal plane D$n$ ($n=1,2,3\ldots$), proportional to $\eta_0$ as in Eq.\ (\ref{eq3}), can be modified accordingly. The seed deformation $\eta_0$ can be tuned by the ejection pressure of the jet. For example, with an ejection pressure increased to 2.0 bar, we obtained a new set of QDM's with deformations of 35\% (D1), 22\% (D2), 14.4\% (D3), 11\% (D4), 7.6\% (D5), etc, increased from 32\% (D1), 19.2\% (D2), 13.3\% (D3), 10.2\% (D4), 6.5\% (D5), etc at 1.8 bar.

The pinching of the nozzle in one direction in the fabrication process has introduced different inner wall slope and this differential wall slope in turn induces different initial radial velocities, which makes it possible to tune the cavity deformation. The initial radial velocity just outside the orifice can be obtained from Eq.\ (\ref{eq1}) by a taking time derivative:
\begin{equation}
v_r(\theta)\approx \left(\frac{2\pi a}{T}\right) \eta_0 \cos \xi  \cos2\theta\;,
\label{eq5}
\end{equation}
where we have neglected the damping effect in taking the time derivative. Due to the guided compression by the inner wall of the nozzle, the radial velocity inside the nozzle is proportional to the ejection velocity. Therefore, we can assume that the
initial radial velocity of the quadrupole oscillation is also proportional to the jet velocity in the $z$ direction just outside
the orifice:
\begin{equation}
v_r(\theta)=K v_z \cos 2\theta\;,
\label{eq6}
\end{equation}
where $K$ is a constant depending only on the inner geometry of the nozzle and the properties of the liquid. It is independent of the ejection pressure. From Eqs.\ (\ref{eq5}) and (\ref{eq6}), we can see that $\cos\xi$ must be proportional to $v_z$:
\begin{equation}
\eta_0 \cos\xi=Kv_z T/2\pi a\;.
\label{eq7}
\end{equation}
With this and the relation $\eta_i=\eta_0\sin\xi$, we then obtain the following result showing that the seed deformation can be controlled by the jet ejection velocity.
\begin{equation}
\eta_{0} = \sqrt{\eta_i^2+\left(\frac{KT}{2\pi a}\right)^{2} v_{z}^2}\;.
\label{eq8}
\end{equation}
For the representative values of the parameters in Eq.\ (\ref{eq8}), $T\sim$ 30 $\mu$s, $v_z\sim$ 10 m/s, $a\sim$ 15
$\mu$m, $\eta_i\sim$ 0.7 and $K\sim$ 0.1, the ratio $(2\pi a\eta_i/Kv_z T)\sim$ 2.2. The jet ejection velocity is easily
controlled by the ejection pressure.

It is noted that the oscillation of the quadrupole deformation along the $z$ direction is analogous to a damped harmonic oscillator with a non-zero launching velocity. The dependence of $\xi$ and consequently the amplitude of deformation oscillation on the jet velocity $v_z$ is illustrated in Fig.\ 2. In addition, the contribution due to the damping in the initial radial velocity is much smaller than that of the harmonic oscillation. For the parameter values in Fig.\ 2, the ratio is $T/2\pi \tau\sim 0.1$.

\begin{figure}
\includegraphics[width=3.4in]{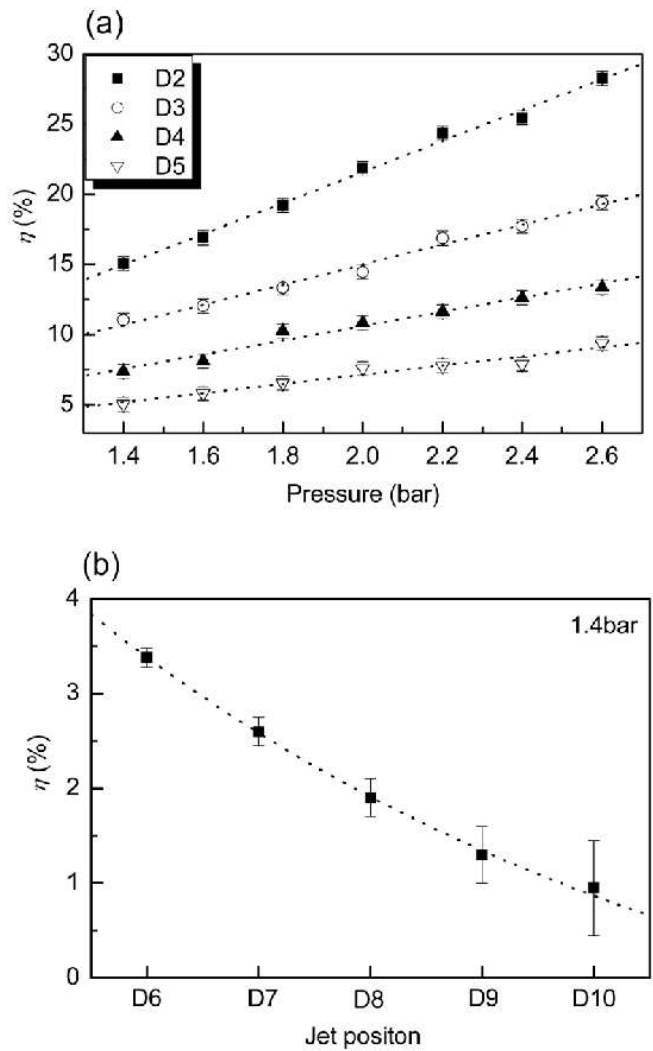}
\caption{(a) Variation of deformation parameter according to the ejection pressure of the jet at  D2, D3, D4 and D5 positions. (b) For deformation below 4\%, the deformation is tunable by selecting different anti-nodal planes at a fixed jet pressure of 1.4 bar.}
\label{fig3}
\end{figure}

Deformation tuning by the jet ejection pressure is experimentally demonstrated in Fig.\ \ref{fig3}, where the degrees of deformation $\eta_n$ at several anti-nodal planes D$n$'s ($n=2,3,4,5$) are plotted as a function of the ejection pressure $P$, which is varied from 1.4 bar to 2.6 bar by a 0.2-bar step. A stable microjet is obtained in this pressure range. By choosing a proper combination of the anti-nodal planes D$n$ and the jet pressure $P$, we can tune the deformation to any value from 5\% (D5, 1.4 bar) to 28\% (D2, 2.6 bar) as shown in Fig.\ 3. Some deformation can be achieved with more than one combination of D$n$ and $P$.

Small deformation from 0\% to 4\% can be obtained around high anti-nodal planes (D6$\sim$D10) under a fixed pressure. The small deformation can be more precisely measured with spectroscopic methods in the next section than with the diffraction patterns since the diffraction patterns for such slightly deformed QDM's are not distinct enough to be used for determining their deformations reliably.

\section{Shape confirmation by spectrum analysis}
\subsection{Method and Principle}
Low-order whispering gallery modes in a circular microcavity reside very close to the cavity boundary. As the microcavity is deformed from a circle to an asymmetric shape while the internal area is preserved, these low-order modes experience wavelength shifts. Since the low-order modes still reside very close to the cavity boundary even in a deformed cavity, these wavelength shifts can be related to the change in the perimeter of the cavity. 

The wavelength shifts due to the change in the perimeter of a great circle of a micro droplet under volume-preserving shape deformation have been investigated before \cite{Tzeng85, Chen93}. In that case, the observed wavelength shift was mostly due to the size increase of the great circle under consideration and thus did not give definitive information on the shape deformation itself for the great circle. In the present work, however, we consider shape deformation for a two-dimensional microcavity with its internal area preserved, and therefore, by measuring the wavelength shift, we can find out how much degree of deformation the cavity has.

Spectra of cavity modes can be observed in several different ways. We use the cavity modified fluorescence of cavity medium. Specifically, we measure the fluorescence of dye molecules resolved in the microjet. Due to the cavity quantum electrodynamics effect, the fluorescence spectrum exhibits enhancement wherever the cavity modes are located and thus it replicates the resonance lines of the cavity modes.

The setup for spectrum measurement is similar to that of Ref.\ \cite{Lee02}. The microjet is excited by an argon-ion pump laser at 514 m. The pump laser is focused with a cylindrical lens into a thin profile with a thickness of 10 micron in the $z$ direction so that a thin slab can be selected to be a two-dimensional microcavity in the microjet. The fluorescence emitted from this region is collected by an objective lens and delivered to a spectrometer with a charge-coupled-device detector. The observed fluorescence spectrum shows several sequences of peaks with well-defined free spectral ranges (FSR's). The modes separated by the same FSR are the modes of the same mode order. In this way, we can identify the mode orders of all of the observed modes.

\begin{figure}
\includegraphics[width= 3.4in]{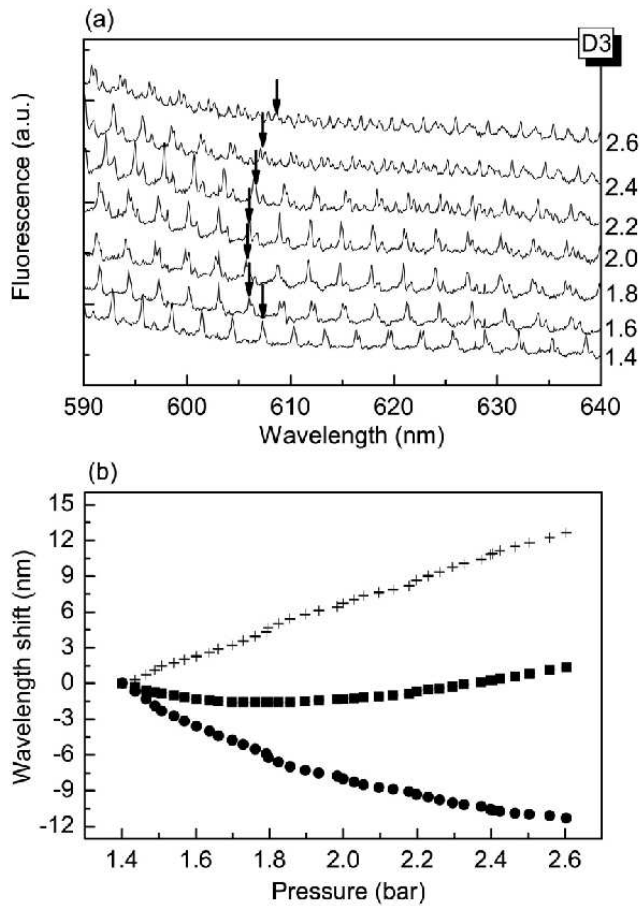}
\caption{(a) Cavity-modified fluorescence spectra observed at D3 as the jet pressure is varied. A particular mode of a good visibility marked by arrows is followed as the jet pressure is changed. (b) Observed wavelength shifts of the particular mode in (a), measured for the jet pressure varied at a small interval, are denoted by filled squares. Wavelength blue shifts due to the area contraction as the pressure is increased are represented by filled circles.  The wavelength shifts due to the deformation only, denoted by crosses, are obtained by subtracting the latter from the former. Error bars are smaller than the symbol sizes.}
\label{fig4}
\end{figure}

A low-order mode, which has a good visibility, can be tracked in the spectrum as the deformation is varied. Some examples are shown in Fig.\ \ref{fig4}(a). Since the low-order mode reside very close to the cavity boundary as discussed above, the resonance wavelengths of these modes are approximately proportional to the cavity perimeter $S$, given by

\begin{equation}
S=\int_{0}^{2\pi} d\theta \sqrt{\left(dr/d\theta\right)^{2}+r^{2}} \;,
\end{equation}
where
\begin{equation}
r(\theta)=\frac{a}{\sqrt{1+\eta^{2}/2}}(1+\eta \cos 2\theta)\;,
\end{equation}
and $\eta$ is the quadrupole deformation parameter, {\em i.e.}, one of $\eta_n$'s at anti-nodal planes. The square root factor in the denominator is a normalization factor needed to make the area of the quadrupole conserved regardless of $\eta$. A solid curve in Fig.\ \ref{fig5} shows the red shift for given deformation according to Eqs.\ (9) and (10). This phenomenon will be refered to as `pure deformation effect' from now on.

\subsection{Experimental results}
Figure  \ref{fig4}(a) shows the observed spectra for various jet pressure values. The peaks marked by arrows correspond to the same mode undergoing wavelength shifts. Anti-nodal plane D3 was chosen and the peak was tracked from 1.4 bar to 2.6 bar. The deformation parameters measured by the diffraction technique were 11.1\%, 12.0\%, 13.3\%, 14.4\%, 16.9\%, 17.7\% and 19.4\%, respectively, in the ascending order of the listed pressure values. In Fig.\ 4(b), the wavelength shifts with respect to the wavelength at 1.4 bar, measured at 0.25 bar interval, are plotted in filled squares. Contrary to our expectation, they show blue shifts as the deformation increases from 1.4 bar to 2.0 bar, beyond which they start to show red shifts. We have found that the blue shift occurs because the cross-sectional area contracts as the ejection pressure, and accordingly the ejection velocity, increases due to the Bernoulli effect.

We checked how much area contraction occurs by measuring spectra at anti-nodal plane D8. Since the deformation there is less than 2\%, the deformation effect can be neglected, and thus the wavelength shift of a particular mode due to area contraction can be measured as a function of the ejection pressure Since area contraction accompanies reduction in perimeter, it leads to a blue shift. Filled circles in Fig.\  \ref{fig4}(b) represent the observed wavelength shifts at D8 as a function of the ejection pressure. They indeed exhibit pure blue shifts. The area contraction is believed to be completed immediately after the jet leaves the orifice.

By subtracting the wavelength shifts due to the area contraction from the overall wavelength shifts, we obtain the wavelength shifts due to the deformation effect only. The results are shown in Fig.\  \ref{fig4}(b) as crosses, exhibiting pure red shifts as expected.

The range of deformation obtained from D3 is limited, from 11\% to 19.4\%. Smaller deformation can be obtained from D4 and D5. Since the measurement of wavelength shift requires that the same mode be followed as the deformation changes, we need to find the same mode in D4 and D5 as the one used in D3. We can directly compare the spectrum obtained from D3 at 1.4 bar with the spectrum from D4 at 2.0 bar since both give 11\% deformation as shown in Fig.\ \ref{fig3}(a). By using the fact that the same mode should have the same FSR, we can identify the mode in the spectrum from D4
that is the same as the mode tracked in the spectrum from D3. Once the same mode is found, we can adjust the pressure and extend the wavelength shift measurement from 2.6 bar down to 1.4 bar for D4. Similarly, we can connect the measurement from D4 to D5. In this way, we can follow one particular mode with a good visibility across several anti-nodal planes and measure the wavelength shift for deformation from 5\%(at 1.4 bar in D5) to 19.4\% (at 2.6 bar
in D3).

\subsection{Small deformation case}
\begin{figure}
\includegraphics[width=3.4in]{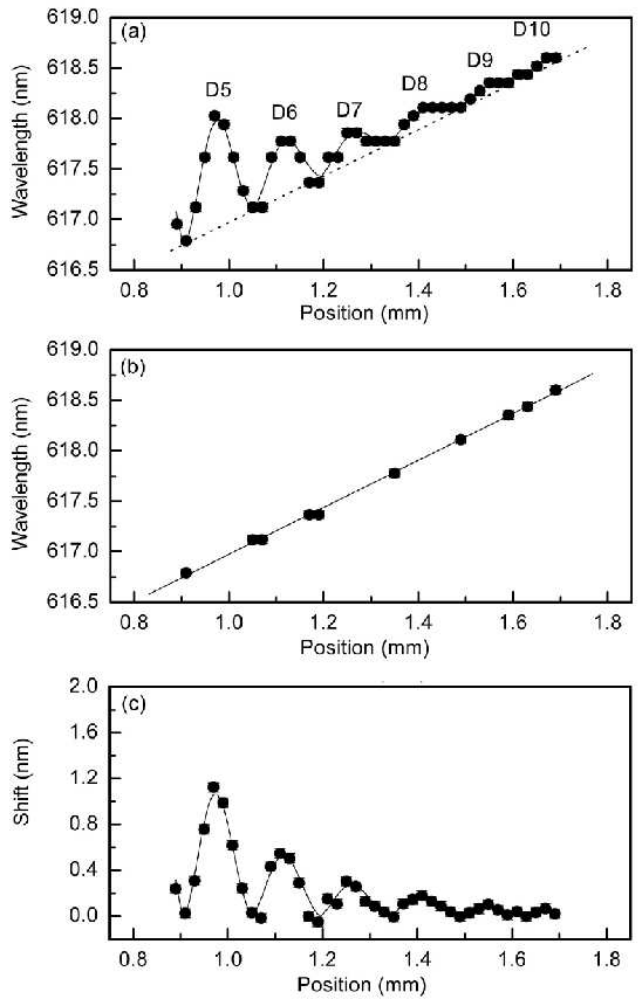}
\caption{(a) Wavelength shifts in the small deformation region. Oscillating pattern shows the deformation effect and the
underlying slope is due to the area expansion as the jet goes up. (b) Wavelength shifts due to the area expansion can be obtained by joining the local minima in (a), corresponding to no deformation. (c) Wavelength shifts due to deformation only is obtained by subtracting the shifts due to the area expansion in (b) from the total shifts in (a). Solid lines are the fits. The fitting parameters in (c) are $\eta_0$=0.7, $\tau=35\; \mu$s, $T=28\; \mu$s, $\xi$=2.1 for $v_z$=10 m/s. Error bars are
smaller than the point sizes. } 
\label{fig6}
\end{figure}

Deformation smaller than 5\% can be obtained from the region of jet from D5 and above. Since the deformation is so small there, the surface profile is nearly flat, and thus any thin slice in this region can serve as a two-dimensional microcavity. We can easily track one mode in the observed spectra while we scan the excitation location continuously with the ejection pressure of the jet fixed. The observed wavelength shifts for various excitation location on the jet at 1.4 bar are shown in Fig.\ \ref{fig6}(a). The oscillatory behavior is due to the oscillation of the cavity perimeter according to the oscillation of the degree of deformation as in Eq.\ (\ref{eq1}). The overall underlying red shift is due to the slight area expansion as the jet goes up. The vertical velocity of the jet $v_z$ decreases slightly as the jet goes up due to air friction and the gravity. Since the flux of the jet should be conserved, the cross sectional area slightly expands as the velocity drops.

The red shift can be easily extracted from the observed shifts. The local minima in the observed shifts must correspond to no deformation ({\em i.e.}, circle) and they are fit well by a straight line as shown in Fig.\ \ref{fig6}(b). These local minimum values are the red shifts due to the area expansion. The amount of red shift over a distance from one anti-nodal plane to the next is only 0.3 nm. After subtracting the amount of red shifts corresponding to this linear fit from the observed shifts, we obtain the shifts due to the deformation only as shown in Fig.\ \ref{fig6}(c).

\begin{figure}
\includegraphics[width=3.4in]{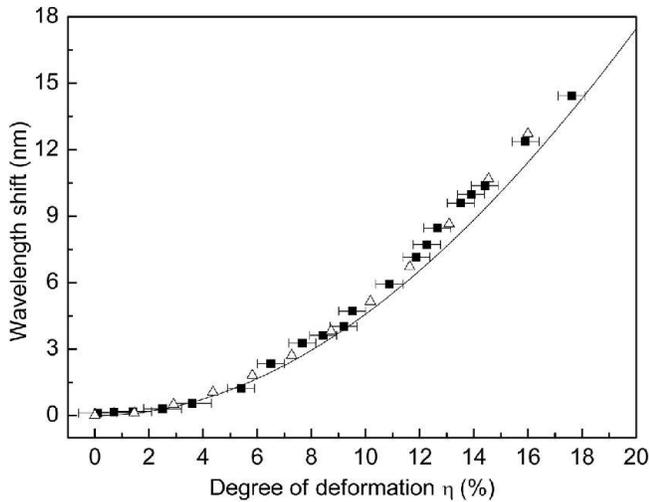}
\caption{Observed wavelength red shifts vs.\ the degree of deformation measured by the diffraction technique. Solid line is
the calculated red shift from the perimeter of a quadrupole, open triangles are the result of wave calculation for the same cavity and filled squares are the experimental results. Dotted line is the calculated red shift for an ellipse with its area conserved and with the same deformation as the quadrupole. Vertical error bars are smaller than the point size.} 
\label{fig5}
\end{figure}

The observed oscillatory behavior conforms to an effective deformation parameter $\eta_{\rm eff}$ given by
\begin{equation}
\eta_{\rm eff}(z)=\eta_{0}\exp\left(-\frac{z}{v_z \tau}\right)\left| \sin\left(\frac{2\pi}{v_{z}T}z+\xi\right)\right|\;.
\label{eq11}
\end{equation}
The zero shift occur when the sine factor vanishes (no deformation) whereas the local maxima occurs when the sine factor
is $\pm 1$ (at the anti-nodal planes). The experimental points for the degree of deformation below 5\% in Fig.\ \ref{fig5} are
obtained from the peak shift values at the anti-nodal planes D6$\sim$D10.

We can calculate the perimeter of a quadrupole defined by Eq.\ (10) and from this perimeter we can calculate wavelength shift due to deformation only with respect to the wavelength for $\eta=0$. The result is shown as a solid curve in Fig.\ \ref{fig5}. In our experiment, we can measure the degree of deformation as a function of the jet pressure by using the diffraction technique as mentioned before. We can also measure the wavelength red shift due to deformation only for a given jet pressure as discussed above. We then obtain a series of pairs of a red shift and a degree of deformation for various jet pressure values. For small deformation, we measure the red shift for a fixed pressure but at different position on the jet as discussed above. The results are shown as filled squares in Fig.\ \ref{fig5}. The experimental results are in good agreement with the calculation assuming quadrupole deformation.

Small deviation observed for large deformation is due to the fact that considering the low order modes followed in spectra as a ray circulating very close to the cavity boundary is just too much simplification. We can solve the wave equation for a given quadrupole cavity with the same size parameter as in the experiment and find resonance wavelengths of the same mode observed in the experiment. Details of mode calculations are given in Ref.\ \cite{JBShim}. The resulting wavelength shifts from the wave calculations are shown in Fig.\ 6 as open triangles, which also show about the same amount of deviation from the solid curve by the simple model. The observed wavelength shifts match well with the results of the wave calculation, and therefore, we conclude that the shape of our microcavity can be regarded as a quadrupolar deformed microcavity within the limit of our experimental error.

This work was supported by National Research Laboratory Grant and by Korea Research Foundation Grant (KRF-2002-070-C00044, -2005-070-C00058). SWK was supported by KOSEF Grant (R01-2005-000-10678-0).

\bibliographystyle{prsty}

\begin{references}

\bibitem{Nockel94}
J.\ U.\ N\"{o}ckel, A.\ D.\ Stone, and R.\ K.\ Chang, Opt. Lett. {\bf 19}, 1693 (1994).

\bibitem{Mekis95}
A.\ Mekis, J.\ U.\ N\"{o}ckel, G.\ Chen, A.\ D.\ Stone, and R.\ K.\ Chang, Phys. Rev. Lett. {\bf 75}, 2682 (1995).


\bibitem{Chang96}
R.\ K.\ Chang and A.\ K.\ Campillo, eds., {\em Optical Processes in Microcavities }(World Scientific, Singapore, 1996)

\bibitem{Nockel97}
J.\ U.\ N\"{o}ckel and A.\ D.\ Stone, Nature {\bf 385}, 45 (1997).

\bibitem{Gmachl98}
C.\ Gmachl, F.\ Capasso, E.\ E.\ Narimanov, J.\ U.\ N\"{o}ckel, A.\ D.\ Stone, G.\ J.\ Faist, D.\ L.\ Sivco, and A.\ Y.\ Cho, Science {\bf 280}, 1493 (1998).

\bibitem{Gianordoli00}
S.\ Gianordoli, L.\ Hvozdara, G.\ Strasser, W.\ Schrenk, J.\ Faist and E.\ Gornik, IEEE J. Quantum Electron. {\bf 36}, 094102 (458).

\bibitem{Gmachl02}
C.\ Gmachl, E.\ E.\ Narimanov, F.\ Capasso, J.\  N.\ Baillargeon, and A.\ Y.\ Cho, Opt. Lett. {\bf 27}, 824 (2002).

\bibitem{Rex02}
N.\ B.\ Rex, H.\ E.\ Tureci, H.\ G.\ L.\ Schwefel, R.\ K.\ Chang, and A.\ D.\ Stone, Phys.\ Rev.\ Lett.\ {\bf 88}, 094102 (2002).

\bibitem{Qian86}
S.\ Qian, R.\ K.\ Chang, Phys.\ Rev.\ Lett.\ {\bf 56}, 926 (1986).

\bibitem{Campillo91}
A.\ J.\ Campillo, J.\ D.\ Eversole, H.\ Lin, Phys.\ Rev.\ Lett.\ {\bf 67}, 437 (1991).

\bibitem{Lee02}
S.\ B .\ Lee, J.\ H.\ Lee, J.\ S.\ Chang, H.\ J.\ Moon, S.\ W.\ Kim, and K.\ An, Phys.\ Rev.\ Lett.\ {\bf 88}, 033903 (2002).

\bibitem{Vernooy98}
D.\ W.\ Vernooy, V.\ S.\ Ilchenko, H.\ Mabuchi, E.\ W.\ Streed, H.\ J.\ Kimble, Opt.\ Lett.\ {\bf 23}, 247 (1998).

\bibitem{Moon2003}
H.-J. Moon and K. An, Jpn.\ J.\ Appl.\ Phys.\ {\bf 42}, 3409 (2003).

\bibitem{Vahala03}
D.\ K.\ Armani, T.\ J.\ Kippenberg, S.\ M.\ Spillane, K.\ J.\ Vahala,  Nature {\bf 421}, 925 (2003).

\bibitem{Lacey03}
S.\ Lacey, H.\ Wang, D.\ H.\ Foster, J.\ U.\ N\"{o}ckel, Phys.\ Rev.\ Lett.\ {\bf 91}, 033902 (2003).

\bibitem{Berglund73}
R.\ N.\ Berglund, B.\ Y.\ H.\ Liu, Environ.\ Sci.\ Technol.\ {\bf 7},
147(1973).

\bibitem{Lin90}
H.\ Lin, J.\ D.\ Eversole, A.\ J.\ Campillo,  Rev.\ Sci.\ Instrum.\ {\bf 61}, 1018 (1990).

\bibitem{Moon95}
H.\ Moon, Guang-Hoon Kim, Yong-Sik Lim, Chun-Soo Go, Jai-Hyung Lee, and Joon-Sung Chang, Rev.\ Sci.\ Instrum.\ {\bf 66}, 3030 (1995).

\bibitem{Lee06}
S.-B .\ Lee, J.\ Yang, S.\ Moon, J.-B.\ Shim, H.-W.\ Lee, S.-W.\ Kim, J.-H.\ Lee, and K.\ An, ``Diffraction patterns by quadrupolar deformed microcavities'', to be published.

\bibitem{Lamb45}
Lamb, {\em Hydrodynamics}, pp.473-475 (Dover, New York, 1945)

\bibitem{Tzeng85}
H.\-M.\ Tzeng, M.\ B.\ Long, R.\ K.\ Chang, and P.\ W.\ Barber, Opt.\ Lett.\ {\bf 10}, 209(1985).

\bibitem{Chen93}
G.\ Chen, M.\ M.\ Mazumder, Y.\ R.\ Chemla, A.\ Serpeng\"{u}zel, and R.\ K.\ Chang Opt.\ Lett.\ {\bf 18}, 1993 (1993).

\bibitem{JBShim}
J.-B.\ Shim and H.-W.\ Lee, S.-B.\ Lee, J.\ Yang, S.\ Moon, J.-H.\ Lee, and K.\ An, S.-W.\ Kim, ``Regular spectra and universal directionality of emitted radiation from a quadruplar deformed microcavity'', arxiv.org/abs/physics/0603221.

\end{references}

\end{document}